\documentclass[10pt,twocolumn]{article}
\usepackage[letterpaper,margin=0.8in]{geometry}
\usepackage{hyperref}
\usepackage{natbib}
\usepackage{astrojournals}
\usepackage{graphicx}
\usepackage{mathtools}
\usepackage{amsmath}
\usepackage{abstract}
\usepackage{lipsum}
\usepackage[charter]{mathdesign}
\usepackage{microtype}
\usepackage{xspace}
\usepackage{xcolor}

\usepackage{caption}
\usepackage{sectsty}
\allsectionsfont{\raggedright\bfseries\large}
\subsectionfont{\raggedright\normalfont\itshape\normalsize}

\newcommand*{\fhelper}[1]{\ensuremath{\vec{f}_{\!\text{#1}}}\xspace}
\newcommand*{\ihelper}[1]{\ensuremath{i_{\!\text{#1}}}\xspace}
\newcommand*{\fd}{\fhelper{d}}
\newcommand*{\fr}{\fhelper{r}}
\newcommand*{\ft}{\fhelper{t}}
\newcommand*{\fj}{\fhelper{j}}
\newcommand*{\ieangle}{\ihelper{e}}

\newcommand{\sgra}{Sgr~A$^*$\xspace}
\newcommand{\eg}{e.\,g.\xspace}

\renewcommand{\deg}{\ensuremath{^{\circ}}\xspace}

\renewcommand*{\vec}[1]{\boldsymbol{#1}}

\let\oldhat\hat
\renewcommand*{\hat}[1]{\vec{\oldhat{#1}}}

\hypersetup{%
 pdftitle={Drag force on G2},
 pdfauthor={\textcopyright\ authors},
 bookmarksopen=true,
 colorlinks=true,
 linkcolor=black,
 citecolor=black,
 urlcolor=black}

\defcitealias{McCourt2016}{MM16}

\usepackage{etoolbox}
\makeatletter

\patchcmd{\NAT@citex}
  {\@citea\NAT@hyper@{\NAT@nmfmt{\NAT@nm}\NAT@date}}
  {\@citea\NAT@nmfmt{\NAT@nm}\NAT@hyper@{\NAT@date}}
  {}
  {}

\patchcmd{\NAT@citex}
  {\@citea\NAT@hyper@{%
     \NAT@nmfmt{\NAT@nm}%
     \hyper@natlinkbreak{\NAT@aysep\NAT@spacechar}{\@citeb\@extra@b@citeb}%
     \NAT@date}}
  {\@citea\NAT@nmfmt{\NAT@nm}%
   \NAT@aysep\NAT@spacechar%
   \NAT@hyper@{\NAT@date}}
  {}
  {}

\patchcmd{\NAT@citex}
  {\@citea\NAT@hyper@{%
     \NAT@nmfmt{\NAT@nm}%
     \hyper@natlinkbreak{\NAT@spacechar\NAT@@open\if*#1*\else#1\NAT@spacechar\fi}%
       {\@citeb\@extra@b@citeb}%
     \NAT@date}}
  {\@citea\NAT@nmfmt{\NAT@nm}%
   \NAT@spacechar\NAT@@open\if*#1*\else#1\NAT@spacechar\fi%
   \NAT@hyper@{\NAT@date}}
  {}
  {}

\makeatother

\begin{document}
\fontsize{10}{14}\selectfont

\title{Using gas clouds to probe the accretion flow around \sgra: \\G2's delayed pericenter passage}

\author{Ann-Marie Madigan%
\thanks{UC Berkeley, Berkeley, CA; ann-marie@astro.berkeley.edu},
Michael McCourt,%
\thanks{UC Santa Barbara, Santa Barbara, CA}
\, \& Ryan O'Leary%
\thanks{JILA, University of Colorado and NIST, Boulder, CO}}
\date{\today}

\twocolumn[
\addtocounter{footnote}{1}     
\maketitle
\begin{onecolabstract}
We study the dynamical evolution of the putative gas clouds G1 and G2 recently discovered in the Galactic center.  
Following earlier studies suggesting that these two clouds are part of a larger gas streamer, we combine their orbits into a single trajectory.  
Since the gas clouds experience a drag force from background gas, this trajectory is not exactly Keplerian.
By assuming the G1 and G2 clouds trace this trajectory, we fit for the drag force they experience and thus extract information about the accretion flow at a distance of thousands of Schwarzschild radii from the black hole.
This range of radii is important for theories of black hole accretion, but is currently unconstrained by observations.

In this paper we extend our previous work by accounting for radial forces due to possible inflow or outflow of the background gas.  
Such radial forces drive precession in the orbital plane, allowing a slightly better fit to the G1 and G2 data.
This precession delays the pericenter passage of G2 by 4\,--\,5 months relative to estimates derived from a Keplerian orbital fit; if it proves possible to identify the pericenter time observationally, this enables an immediate test of whether G1 and G2 are gas clouds part of a larger gas streamer.
If G2 is indeed a gas cloud, its closest approach likely occurred in late summer 2014, after many of the observing campaigns monitoring G2's anticipated pericenter passage ended.
We discuss how this affects interpretation of the G2 observations.
\end{onecolabstract}
\vspace*{2\baselineskip}
]
\saythanks

\raggedbottom
\section{Introduction}
\label{sec:intro}
\begin{figure*}
  \centering
  \includegraphics[width=\textwidth]{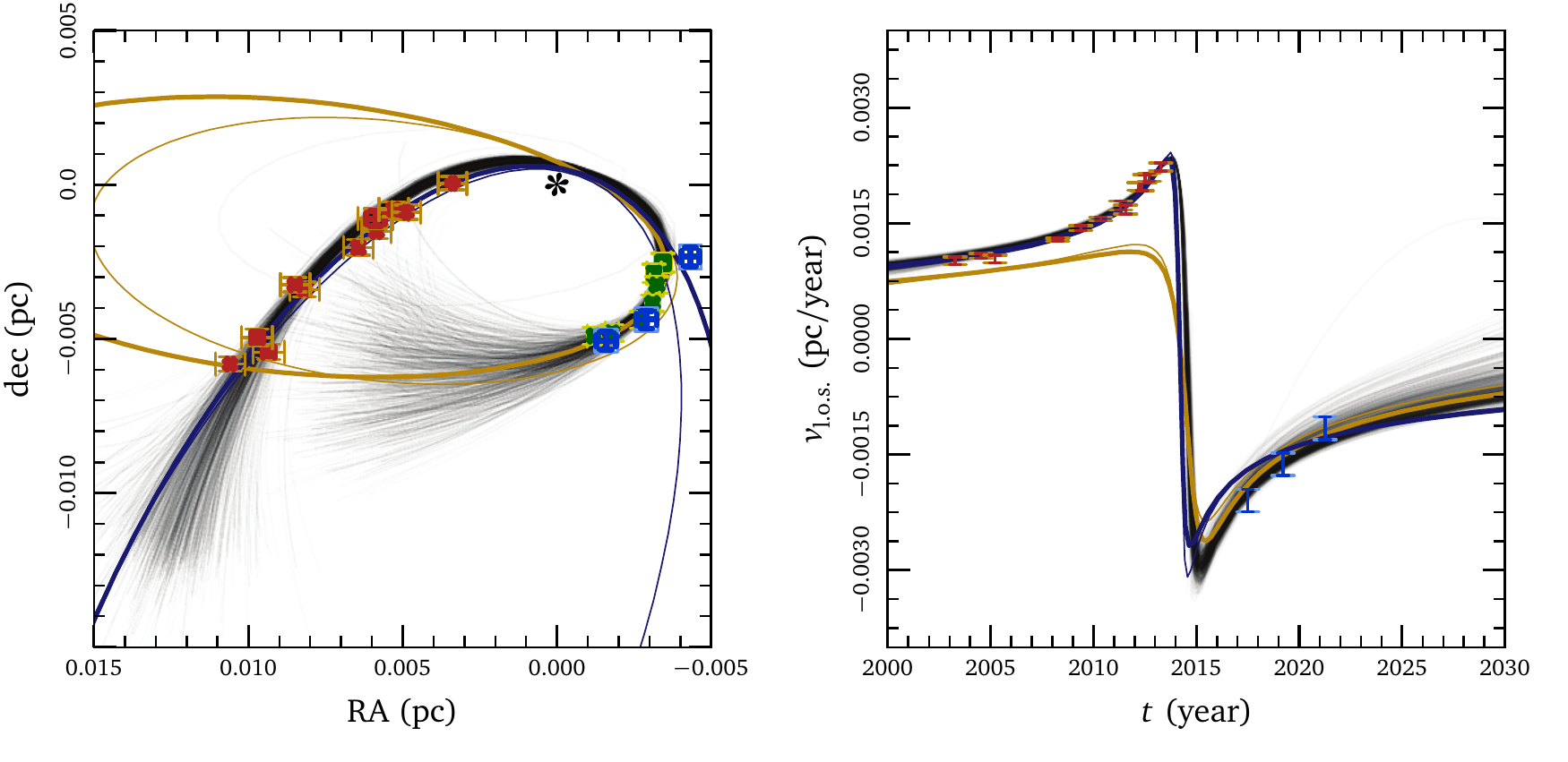}
  \caption{(\textit{left:}) Comparison of our model results with astrometric data.
Thin black lines show a random sample of models drawn from our MCMC chain.
Red points indicate Br-$\gamma$ observations of G2 from \citet{Gillessen2013b}, while the green and blue points show L-band and Br-$\gamma$ observations of G1 from \citet{Pfuhl2015}. 
Smaller, dark error bars correspond to those reported in the observational papers. 
Larger error bars include the systematic error found from our maximum-likelihood analysis (see~\citetalias{McCourt2016} for details).
Colored ellipses show Keplerian orbits fit to G1 (yellow) and G2 (blue).
Thick lines show the fits we derived in \citetalias{McCourt2016}, and thin lines show the fits described in \citet{Gillessen2013b} and in \citet{Pfuhl2015}; the differences between them provides a measure of the uncertainty in the fitting.
The location of Sgr~A$^{*}$ is marked with an asterisk. 
(\textit{right:}) Comparison of our models with line-of-sight velocity.
As in the astrometry plot, the smaller error bars are from \citet{Gillessen2013b} and \citet{Pfuhl2015}, while the larger ones account for the systematic error determined by our maximum-likelihood analysis. 
The G1 data points (blue error bars) and Kepler fits (yellow curves) are shifted by 13 years to compare with our models \citep[cf.][]{Pfuhl2015}.}\label{fig:orbit-bundle}
\end{figure*}
\sgra, the black hole in the center of our Galaxy, is our closest
example of an accreting super-massive black hole.  Even in this nearby
example, important aspects about how gas makes its way to the black hole
remain unknown. 
For example, the accretion rate at $\sim100~R_s$\footnote{$R_s \equiv
2GM_{\bullet}/c^2$ is the Schwarzschild radius of a non-spinning super-massive
black hole of mass $M_{\bullet}$. For a mass of $M_{\bullet} \sim 4.3 \times 10^6 M_{\odot}$
\citep{Gillessen2009}, $R_s \sim 4 \times 10^{-7}$pc.} (inferred from radio observations) is less than
$\sim1\%$ of the
accretion rate at the Bondi radius ($\sim10^5~R_s$) inferred from X-ray observations
\citep{Quataert2000b}.
Thus, only a tiny fraction of the gas bound to
the black hole makes its way to the event horizon.

Several models have been proposed to explain this reduction in accretion rate.
Examples include ADIOS models, which expel most of the gas in a strong outflow \citep{Blandford1999}, CDAF models, which recycle gas outward in large-scale convection cells while the gas remains gravitationally bound \citep{Quataert2000}, and magnetically arrested models (MAD), in which magnetic forces inhibit radial inflow \citep[\eg][]{Narayan2003}. 
Tests at radii intermediate between the event horizon and the Bondi radius, $\sim\,10^3$--$10^4~R_s$, are crucial to distinguish among these various accretion theories, yet there are few probes of this region.
Recent infrared observations have identified low-mass gas clouds, G1 and G2, moving through this exact region.
Measuring their interaction with the background gas could therefore provide important information about black hole accretion physics \citep[e.g.,][]{Narayan2012}.

Such interaction could take the form of a drag force, which would modify the orbits of the clouds and cause deviations from a Keplerian trajectory.
These deviations are unfortunately too small to unambiguously identify in the published data for the individual clouds, which span only a few years.
Since G1 and G2 are on strikingly similar, though slightly different, orbits about the black hole \citep{Gillessen2012, Phifer2013, Pfuhl2015}, we instead work with the assumption that G1 and G2 are part of a larger gas streamer \citep[from, \eg, a partial tidal disruption;][]{Guillochon2014}, and therefore trace different parts of the same trajectory.

We therefore assume that G1 represents the future evolution of G2; this assumption makes specific predictions, testable on a $\sim$\,3--5 year timescale, or possibly (as we discuss in this paper) with data that's currently available.

First proposed by \citet{Pfuhl2015}, this assumption that G1 and G2 follow the same trajectory effectively extends the length of observations by $\sim$\,13 years and significantly expands our ability to constrain its long-term evolution.
In \cite{McCourt2016} \citepalias[hereafter][]{McCourt2016}, we used this assumption to constrain the rotation axis of the accretion flow surrounding \sgra. 
We found a rotation axis similar to that of the Galaxy; this is consistent with the circumnuclear disk, with the Fermi bubbles, with a possible X-ray jet in the Galactic center, and with available constraints from the Event Horizon Telescope (see MM16 for references and details).

This paper builds on the analysis presented in MM16 by allowing for possible inflow or outflow in the background accretion flow.  We show that the data favor inflow along the trajectory, which drives precession of the eccentricity vector and provides a slightly better fit to the data.  This effect is equivalent to delaying G2's pericenter passage by $\sim5$ months from predictions based on a Keplerian fit to the orbit \citep[e.g.,][]{Phifer2013,Gillessen2013b}.

\section{Method}
\label{sec:method}
As discussed in the introduction, we assume that G1 and G2 are part of a coherent gas streamer, with G1 preceding G2 by a time-lag of about $\sim$\,13\,years (which we fit for).
We therefore assume the two clouds follow a single trajectory. 
This trajectory would be a Kepler orbit if the clouds moved through a vacuum, but deviates due to drag forces from the ambient gas in the accretion flow.
We use the inferred changes in their orbits to probe the accretion flow around the super-massive black hole, \sgra.
This analysis is discussed in \citet{Pfuhl2015} and again in \citetalias{McCourt2016}; we do not describe it in detail here.

We use a simple model for the rotating accretion flow, with the goal of being as agnostic as possible about its structure. 
We define this model by equation\,5 in \citetalias{McCourt2016}.
The model is specified by a single power-law density profile, with exponent $\alpha$, by the rotation parameter $f_\text{kep} \equiv v_\text{rot}/v_\text{kep}$ (assumed constant with radius), by the angular coordinates of the angular momentum vector $\hat{j}$, and by the magnetic field strength, $\beta$.
As in \citet{Pfuhl2015} and in \citetalias{McCourt2016}, we make the approximation that G1 and G2 have the same mass, size, and shape, and that these quantities do not evolve with time.
We have updated our model to include inflow or outflow by adding a radial component to the background velocity (equation 5c in \citetalias{McCourt2016}).

We numerically integrate trajectories for a gas cloud starting from an initial condition, varying the parameters in the model for the accretion flow.
We estimate the likelihood of each trajectory by comparing it against the astrometry and velocity data published in \cite{Gillessen2013a,Gillessen2013b} and in \citet{Pfuhl2015}.
We maximize this likelihood over all free parameters for the accretion flow, using the \texttt{emcee} MCMC optimizer \citep{foreman2013}.
Our methodology is described in detail in \citetalias{McCourt2016} and our code is publicly available online.\footnote{https://github.com/mkmcc/g2-drag-force}

\section{Results}
\label{sec:results}
We compare our model results with astrometric and velocity data in figure~\ref{fig:orbit-bundle}. 
Due to uncertainty in the background model, we obtain a family of degenerate solutions which all fit the data, shown as a ``bundle'' of thin black lines (we show the parameter distributions from our MCMC chain in the appendix).
Our predictions would become far more specific if we were to adopt a particular accretion model, e.\,g. an ADAF \citep{Narayan1994}.
We do obtain a better fit to the data than \citetalias{McCourt2016}: the $\chi^2$ per degree of freedom (d.o.f.) improves to $\chi^2$/d.o.f $ = \{3.1, 0.82\}$ from $\{4.1, 2.8\}$ when we allow for inflow in our model of the background medium.
In both cases, the first number is a standard $\chi^2$ calculated using the uncertainties quoted in \citet{Gillessen2013a}, \citet{Gillessen2013b}, and \citet{Pfuhl2015}, and the second number is calculated including an additional systematic uncertainty derived self-consistently from the data (see equation 7 in \citetalias{McCourt2016}). 

\begin{figure}
  \centering
  \includegraphics[width=\linewidth]{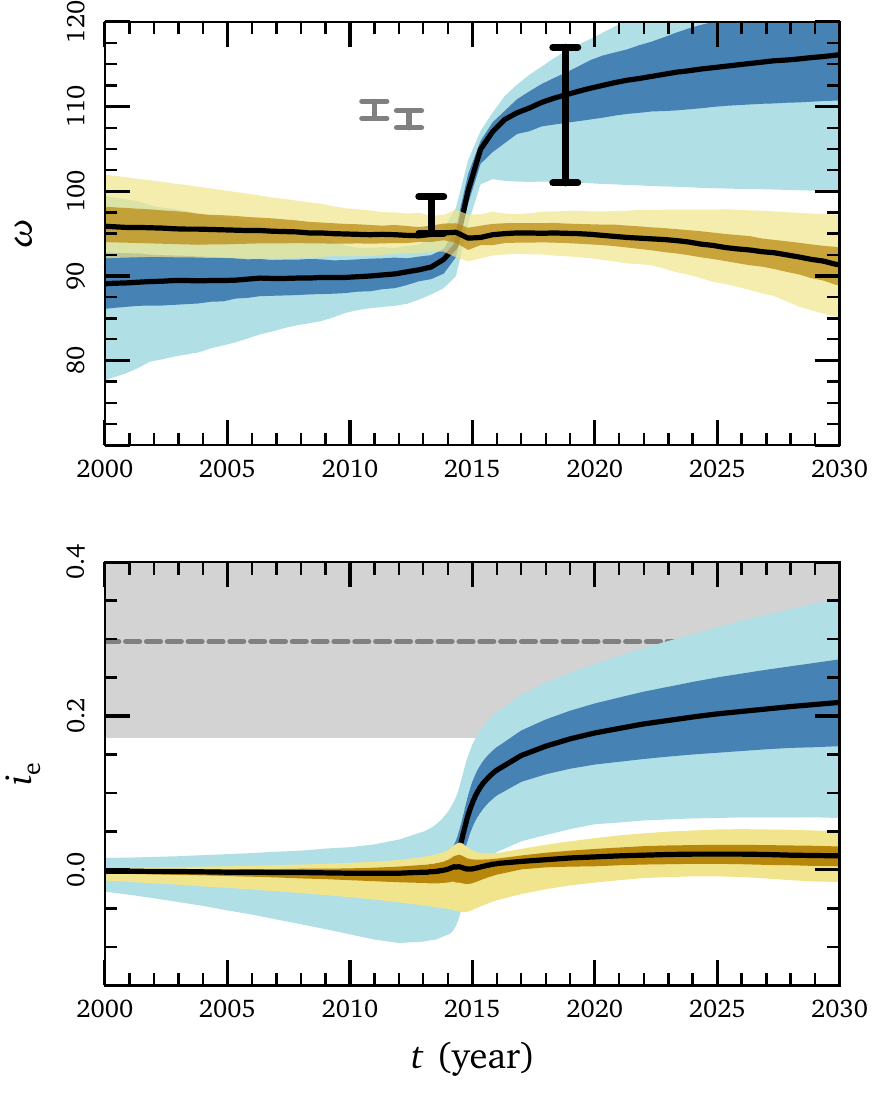}
  \caption{Precession in models with (blue) and without (yellow) inflow.
Shaded regions indicate 1- and 2-$\sigma$ quantiles.
(\textit{top}): argument of pericenter $\omega$ for the osculating orbit.  Over the 30 years shown, its evolution is strongly peaked at pericenter.
(\textit{bottom}): Integrated effect of precession expressed as the rotation angle $i_{\text{e}}$ in radians (see footnote\,3, or \citealt{Madigan2016} for the definition).
The gray band shows a crude estimate derived from Kepler fits to G1 and G2 in \cite{Pfuhl2015} and in \citet{Gillessen2013b}.
Models from \citetalias{McCourt2016} (yellow) do not reproduce this precession.}\label{fig:precession}
\end{figure}
Figure~\ref{fig:precession} compares the eccentricity vector precession in our simulations with the G1 and G2 data.
This can be seen by either the argument of pericenter $\omega$ (\textit{top panel}) or more directly by the rotation angle \ieangle (\textit{bottom panel}), defined in \citet{Madigan2016}.\footnote{For a Keplerian orbit, the eccentricity vector $\vec{e}$ points toward pericenter.  Precession amounts to a rotation of the $\vec{e}$ vector in the plane of the orbit.  We define the direction for prograde precession as $\hat{b}\equiv\hat{j}\times\hat{e}$.  The precession rate $\ieangle^{\prime}$ is defined to be $\hat{b}\cdot\hat{e}^{\prime}$, and $\ieangle \equiv \int \ieangle^{\prime} dt$.}
Shaded regions show the 1- and 2-$\sigma$ spread in our models; yellow curves show results from \citetalias{McCourt2016} which do not include inflow or outflow, blue curves show results from our updated model.  
Models without inflow have almost no net precession; the implied precession derived from Kepler fits to G1 and G2 data (black error bars, or grey band) is much greater than can be accommodated in these models.
Models with inflow can easily reproduce this precession, however (see Appendix A for a discussion connecting precession to inflow). We note that this is a crude comparison to the data, since our model provides a better fit to the data than the separate Kepler fits used to infer observational values for $\omega$ or for \ieangle; figure~\ref{fig:precession} nonetheless illustrates the key difference between our models and the models from \citetalias{McCourt2016}, which did not allow for inflow or outflow in the background: we find that inflow drives prograde precession of the orbit, thus enabling a better fit to the data.

\begin{figure}
  \centering
  \includegraphics[width=\linewidth]{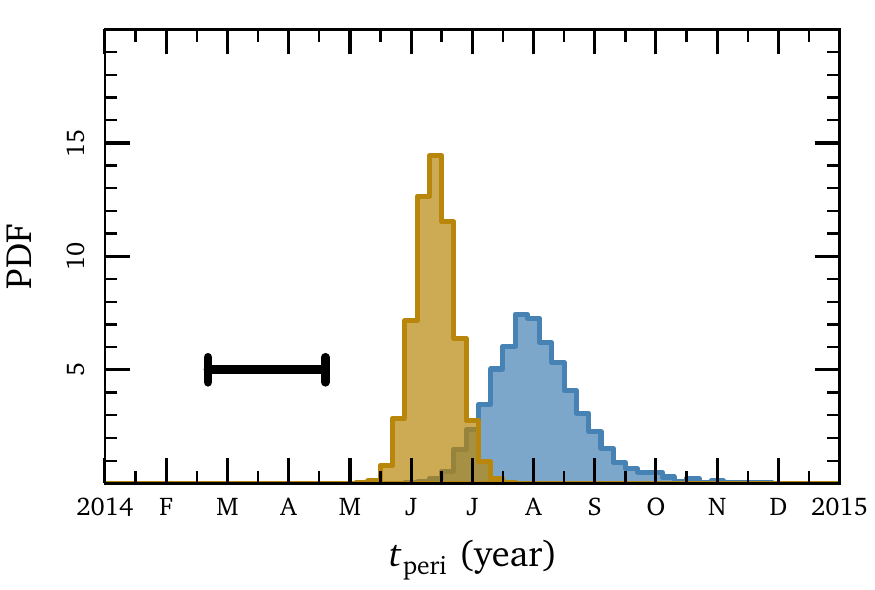}
  \caption{Pericenter times derived for G2.  The prograde eccentricity vector precession seen in models with inflow (blue) implies a later pericenter time than models without inflow (yellow), as is expected.  The error bar shows the pericenter time derived for a Kepler orbit, which is earlier by many months.
  }\label{fig:tperi}
\end{figure}
Physically, prograde precession implies a delay in the pericenter passage of the orbit.
Figure~\ref{fig:tperi} shows this explicitly, and demonstrates that the predicted pericenter time of G2's orbit is model-dependent (even among models which fit the
available data). 
Models with different precession rates can move the pericenter time by many months; this is far greater than the statistical uncertainty from fitting the data with any given model. 
Fitting a purely Keplerian orbit to the G2 data yields the earliest estimate of G2's pericenter passage in March--April 2014\footnote{Note that due to its finite size, G2 should take about two months either side of this time to completely pass pericenter.} \citep{Gillessen2013b}.
The distribution shifts by several months if we include a drag force, peaking mid-June 2014 for models with no inflow, and early August 2014 for models with inflow.

\citet{Witzel2014} report four new epochs of near-infrared imaging of G2 taken from March through early-August 2014.
They constrain the size of G2 in $L^{\prime}$-band in each epoch, finding that it remains compact through the summer of 2014.
Based on the pericenter time expected for a Keplerian orbit, they conclude that G2 has survived its closest approach to \sgra.
We note however, that given the precession implied by comparing G1 and G2, the data in \citet{Witzel2014} could very well have been entirely taken pre-pericenter.
\textit{If it proves possible to identify the pericenter passage observationally to within $\sim$\,2\,months, that could definitively determine whether G2 is an $\sim$\,Earth-mass cloud or a stellar-mass source.}
In principle, this test should be possible using existing data.\footnote{Since G2's orbit is not exactly aligned with Earth's line-of-sight to the galactic center, we note that the time of G2's closest approach is not the same as the time when its line-of-sight velocity changes sign.}

We note that G1's pericenter time is also model-depen\-dent. However, the statistical uncertainty due to the limited data dwarfs the systematic differences between models.
All models are consistent with G1 passing pericenter at essentially any time in 2001.

We find that our models generically prefer inflow (as opposed to outflow) of the background gas; inflow causes prograde precession near pericenter which better fits the orbits of G1 and G2 (we discuss how inflow and outflow influence orbits in the appendix).
If most of the gas in the accretion flow ultimately moves outward in a wind \citep[as is likely to be the case; ][]{Narayan1994,Blandford1999,Quataert2004,Wang2013}, this outflow cannot cover the entire range of angles.
Since G2's orbit is misaligned with the rotation axis of the background gas, we find tentative evidence for a geometrically thick disk with an opening angle of at least 30--50 degrees, broadly consistent with a radiatively inefficient accretion flow.
This is only a tentative result at the moment, however: the magnitude of the inflow is degenerate with other parameters and the effect may be model-dependent.
It is also possible that a different model, with a non-keplerian rotation profile \citep[\eg][]{Pen2003}, could match the observational data without inflow.
This result nonetheless demonstrates the potential power of using the G1 and G2 clouds to probe the Galactic center accretion flow.

\section{Discussion}
\label{sec:discussion}
The G1 and G2 clouds are on strikingly similar, but slightly different orbits around the super-massive black hole in our Galactic center. We interpret these clouds as part of a larger gas streamer; the slight differences in their orbital parameters therefore represent evolution of the trajectory due to interaction with the background gas \citep{Pfuhl2015,McCourt2016}.
If this picture proves correct, the ``G-clouds'' can be used to probe the accretion flow feeding \sgra at distances of 1000s of Schwarzschild radii, a critical range for accretion physics which is currently unconstrained by observations.

In this paper we build upon the work in \citetalias{McCourt2016}, updating our accretion flow model to include a radial component due to inflow or outflow.  
Near pericenter, this radial component drives rapid precession of the orbit, effectively delaying the pericenter time by up to $\sim5$ months relative to predictions based on a Keplerian orbit (see figure~\ref{fig:tperi}).
As in \citetalias{McCourt2016}, we find that we can only fit the data for G1 and G2 if the clouds are elongated and if the background accretion flow rotates.
Uncertainties in the mass and shape of the cloud prevent a determination of the exact properties of the accretion flow;
instead we obtain a degenerate family of models which all fit the data.  
However, more data or a more detailed analysis may distinguish between different models.  
We derive a rotation axis close to both the Galaxy's rotation axis and to the circumnuclear disk.
This axis is consistent with that found in \citetalias{McCourt2016}.

We find that models with inflow better match the data; inflow causes prograde precession of the eccentricity vector near pericenter which is implied by comparing the orbits of G1 and G2, $(\vec{e}_{\text G1} - \vec{e}_{\text G2})\cdot \vec{v}_{\text G2} > 0$ (see Appendix~\ref{subsec:forces} for definitions and details).   
The magnitude of the inflow is degenerate with several other parameters however.  
We find tentative evidence for a geometrically thick accretion disk with an opening angle of at least 30--50 degrees, broadly consistent with radiatively inefficient accretion flow models.
There are a number of observations that can test our model and our assumptions. 
\begin{enumerate}
\item The pericenter time for G2 should be delayed by $\sim$\,five months behind the prediction based on a Kepler orbit.
Though the pericenter passage may be difficult to discern precisely from the data, such a test has the advantage that it could be performed immediately, and with existing datasets.

\item The post-pericenter evolution of G2's orbit should look like G1's in the coming $\sim$3\,--\,10 years (though differences in mass, shape and size of the two gas clouds may translate to a slightly different evolution).\footnote{G1's passage through the accretion flow is likely modifying the flow as they are of similar mass at these radii. This will change the drag force experienced by G2 in its wake.}

\item G1's orbit should continue to circularize and reorient to align with the accretion flow axis. The timescale for this is likely to be decades, however, unless G1 expands
significantly. 

\item ``Older'' clouds (those preceding G1, which have interacted with the accretion flow for a longer time), if they exist, should be less eccentric, have smaller semi-major axes and have angular momentum vectors increasingly aligned with the accretion flow. 
\end{enumerate}

A number of observing campaigns monitored the Galactic center during February--May of 2014, when the cloud was predicted to pass pericenter.
If G2 is in fact a gas cloud, however, its pericenter passage was likely several months afterwards, in late summer of 2014.
Many observing campaigns unfortunately ended before this date. 

\citet{Ponti2015} present x-ray flaring activity of \sgra~ using \textit{Chandra} and \textit{XMM-Newton}.
They report no variation in the flaring rate during the spring 2014.
Towards the end of the campaign however, they observed a series of five bright flares starting in late summer 2014, coinciding with a bright flare detected by \textit{Swift} in September 2014 \citep{Degenaar2015}.
Though these authors conclude the flares are not directly associated with G2's pericenter passage, we note that they coincide closely with the pericenter time implied by our model, shown in figure~\ref{fig:tperi}.

G1's pericenter passage is not well-constrained by the data; it could have happened at any time in 2001.
Unfortunately, \sgra was sparsely monitored around this time. 

If the G-clouds are indeed $\sim$Earth-mass gas clouds, they are one of the only probes of the Galactic center accretion flow at the critical range of radii of thousands of Schwarzschild radii.  
Continued monitoring of the G-clouds will yield constraints on the accretion flow surrounding our nearest super-massive black hole, and provide important insights into the
unknown physics of low-luminosity AGNs in general.

\section*{Acknowledgments}

\noindent{} AM thanks Breann Sitarski and the UCLA Galactic Center Group for insightful discussions during their workshop in December 2015.
AM was supported by UC Berkeley's Theoretical Astrophysics Center.
MM was supported by NASA grant NNX15AK81G.
RMO acknowledges the support provided by NSF grant AST-1313021.
Resources were provided by the NASA High-End Computing (HEC) Program through the NASA Advanced Supercomputing (NAS) Division at Ames Research Center under grant SMD-15-6582.

\bibliographystyle{mn2e}
\bibliography{drag-force-inflow}

\appendix
\section{Observational Implications of Non-Keplerian Forces}
\label{subsec:forces}
A Kepler orbit about a central massive object $M_{\bullet}$ is defined by two vectors: the angular momentum vector ${\vec{j}} = \vec{r} \times \vec{v}$, which defines the orbital plane, and the eccentricity vector $\vec{e} = (\vec{v} \times \vec{j})/ (G M_{\bullet}) - \hat{r}$, which points toward pericenter and orients the orbit within its orbital plane.
The magnitude of the eccentricity vector $\vec{e}$ is the eccentricity $e$ of the orbit; together with the angular momentum $j$, this fixes the energy of the orbiting body.
Both $\vec{j}$ and $\vec{e}$ remain constant in a Kepler potential; changes to these vectors therefore directly probe non-Keplerian forces, which we refer to here as a ``drag'' force \fd.
Time evolution of the $\vec{j}$ and $\vec{e}$ vectors follows:
\begin{subequations}
\begin{align}
{\vec{j}}^\prime &= \vec{\tau}\label{eq:1} \\
{\vec{e}}^\prime &= \frac{\fd \times \vec{j}}{G M_{\bullet}} + \frac{\vec{v} \times \vec{\tau}}{G M_{\bullet}}, & & \label{eq:2}
\end{align}
\end{subequations}
where $\vec{\tau} \equiv \vec{r} \times \fd$ is the torque produced by the drag force.

We can split the drag force into three components:
\begin{equation}
\fd = f_r \hat{\vec{r}} + f_t \hat{\vec{t}} + f_j \hat{\vec{j}}
\end{equation}
The unit vectors $\hat{\vec{r}}$, $\hat{\vec{t}}$ and $\hat{\vec{j}}$ point along the radial direction, tangent to the radial direction in the orbital plane, and along the angular momentum vector of the orbit.  

When applied impulsively at pericenter, each component of the force influences ${\vec{j}}$ and ${\vec{e}}$ in a predictable way:
\begin{enumerate}
\item A tangential force \ft (i.e., aligned with or against the direction of motion) torques the orbit parallel to its angular momentum; this changes the energy and angular momentum (or eccentricity) of the orbit, but not its orientation.  
\item A component of the force out of the orbital plane \fj produces a torque in the orbital plane; this rotates and re-orients the orbit.
This changes the directions of the orbit vectors $\vec{j}$ and $\vec{e}$, but not their magnitudes.
\item Finally, a radial force \fr at pericenter (in addition to the Keplerian force $-GM_{\bullet}\vec{r}/r^3$) drives precession of the eccentricity vector in the orbital plane.
Such a force might result from a non-point-mass contribution to the gravitational potential (from, \eg, a stellar cusp), or from net inflow or outflow of a gaseous background.
\end{enumerate}

Since the effects of drag forces are sharply concentrated near pericenter, \citet{Pfuhl2015} primarily constrained the tangential component of the force \ft.
\citetalias{McCourt2016} included out-of-plane forces \fj by allowing the background to rotate, and found a slightly better fit to the data.
In this paper, we add a purely radial component of the drag force \fr, as might result from a net inflow or outflow of the background; this completes all three possible components of the drag force.

\subsection{Drag Force due to Accretion Flow}
\label{subsec:comps}
In \citetalias{McCourt2016}, we studied how the drag force produced by a rotating accretion flow might influence the $\vec{j}$ and $\vec{e}$ vectors.
We focused on the parallel and perpendicular components of the torque:
\begin{subequations}
\begin{align}
  \tau_{||} &\equiv \vec{\tau}\cdot\hat{j} & \propto \ft \\
  \vec{\tau}_{\perp} &\equiv \vec{\tau} - \tau_{||}\hat{j} & \propto \fj
\end{align}
\end{subequations}
which are observable as a change in the angular momentum of the orbit and from a rotation of its orbital plane, respectively.

In this study, we instead focus on a possible radial component of the force, \fr.  This drives precession, or a rotation of the $\vec{e}$ vector within the orbital plane.
We define a direction for precession as $\hat{b}$, with $\hat{b}\equiv \hat{j}\times\hat{e}$.  Equation~\ref{eq:2} thus shows that the precession rate is given by:
\begin{align}
  \hat{b}\cdot\vec{e}^{\prime} = - \frac{j f_{\text{r}}}{G M_{\bullet}} \cos \psi
  + \frac{\tau_{||} v_{\text{r}}}{G M_{\bullet}} \left[\frac{2}{e} + \cos \psi \right], \label{eq:eb}
\end{align}
where $\psi$ is the true anomaly (or polar angle) of the orbiting body.  In our application to the G2 cloud, the values $f_{\text{r}}$ and $\tau_{||}$ are largest at pericenter. 

The first term in equation~\ref{eq:eb} describes precession of the eccentricity vector $\vec{e}$ due to a radial, non-Keplerian force \fr.
Though this term oscillates in sign over an orbital period ($\propto\cos\psi$), it does not average away because it is sharply peaked at pericenter where $\cos\psi=1$.
The second term in equation~\ref{eq:eb} describes an oscillation of the eccentricity vector $\vec{e}$ over the orbital period due to the perpendicular component of torque $\tau_{||}$.
This term results from fitting Kepler elements to a slightly non-Keplerian orbit; the osculating Kepler ellipse rocks slightly back and forth along the true trajectory.
Though this term averages away over long timescales\footnote{This is true in the limit of a Kepler orbit with constant $\tau_{||}$, not for an ellipse which evolves on a timescale shorter than its orbital period.}, it may dominate the first term at any one time, especially for lower eccentricities.  
As $v_{\text r} \propto \sin\psi$, this term reverses sign at pericenter and at apocenter, a distinctly different signature from the first \fr term.

\begin{figure}
  \centering
  \includegraphics[width=\linewidth]{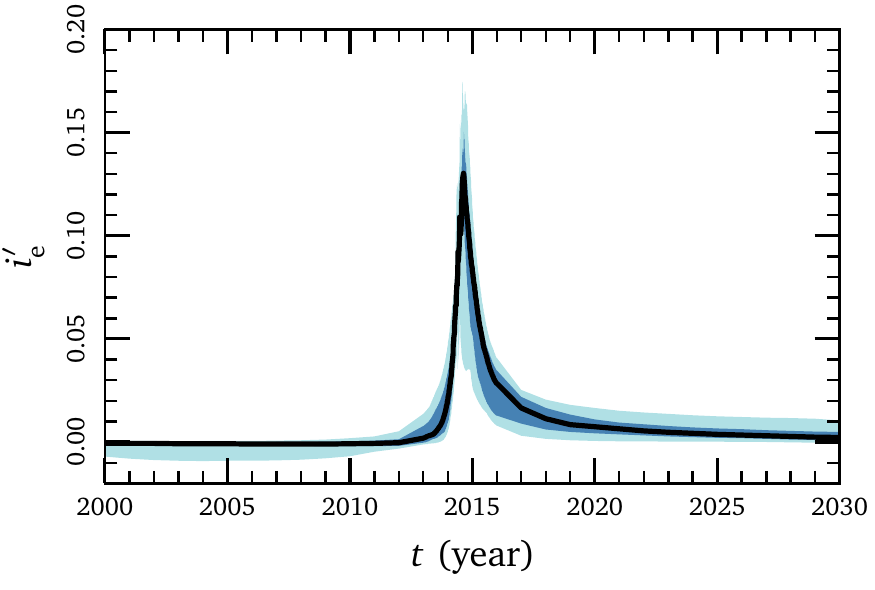}
  \caption{
  Eccentricity vector precession rate $i_{\text{e}}^{\prime}=\hat{b}\cdot\hat{e}^{\prime}$ in radians/year (see equation~\ref{eq:eb}) for models with inflow. It is strongly peaked at pericenter, justifying the impulse approximation described here. 
Shaded regions indicate 1- and 2-$\sigma$ quantiles.
}\label{fig:precession-app}
\end{figure}

In figure~\ref{fig:precession-app} we show that eccentricity vector precession rate is very strongly peaked in the $\sim$year surrounding pericenter, justifying the impulse approximation discussed here. 

The orbit vectors $\vec{j}$ and $\vec{e}$ are constant in a Keplerian potential, but evolve in response to a non-Keplerian force.
Measuring time evolution in the magnitude or direction of $\vec{j}$ and $\vec{e}$ thus directly relates to components of the non-Keplerian force.

\section{Accretion Flow Parameters}
\label{sec:mcmc-chain}
\begin{figure*}
  \centering
  \includegraphics[width=\textwidth]{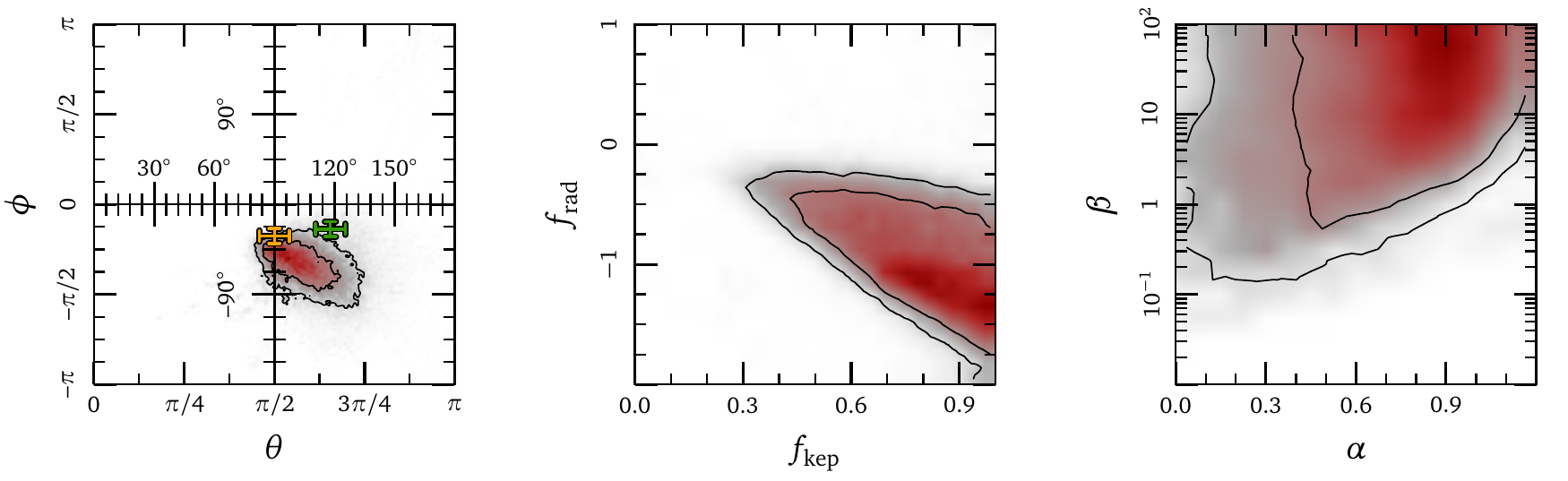}
  \caption[width=\textwidth]{Probability distribution functions of the parameters in our model, determined using a maximum-likelihood analysis.
Lines indicate 1- and 2-$\sigma$ quantiles from the best fit solution.
Left: the rotation axis of the accretion flow in polar angle, $\theta$, and azimuthal angle, $\phi$.
As a comparison, we overplot the rotation axis of the Galaxy (orange) and that of the circumnuclear disk (green).
Middle: the parameters $f_{\text{rad}}$ and $f_{\text{kep}}$ which describe the strength of the inflow and rotation velocity of the accretion flow relative to the local circular Kepler velocity. 
Negative $f_{\text{rad}}$ indicates an inflow solution. 
Right: the slope of the density profile of the accretion flow, $\alpha$, and the magnetic $\beta$ of the plasma, are strongly degenerate with respect to one another as both control the magnitude of the drag force.}
  \label{fig:params}
\end{figure*}

Figure \ref{fig:params} shows the probability distributions for the parameters in our MCMC model.
Our results are as follows:
\begin{enumerate}
  \item The rotation axis of the accretion flow, described in polar ($\theta$) and azimuthal ($\phi$) coordinates\footnote{The angular momentum vector of the accretion flow is defined as  $\hat{j} = \{\sin{\theta} \cos{\phi}, \sin{\theta} \sin{\phi}, \cos{\theta}\}$.}, is essentially unchanged from what we found in \citetalias{McCourt2016}, $\theta \sim (105 \pm 16) \deg$ and $\phi \sim (-51 \pm 23)\deg$. 
In Kepler coordinates \citep[as defined in][]{Lu2009}, our best fitting accretion flow axis corresponds to $(i,\Omega) = (75 \deg\pm 16\deg, 51\deg \pm 23\deg)$.
 For reference, we show the Galactic rotation axis in orange and the circumnuclear disk axis in green.
This result depends on the perpendicular torque $\vec{\tau}_{\perp}$, which results from forces out of the orbital plane.
This prediction is therefore relatively insensitive to radial inflow or outflow in the model.
  \item We find a marginal detection of inflow of the accretion flow, with $f_{\text{rad}} < 0$.
With our simplified model for the background, the data do not allow outflow along the orbit trajectory.
The magnitude of the inflow parameter, $|f_{\text{rad}}|$, is degenerate with the size and shape of the cloud.
The rotation parameter $f_{\text{kep}}$ approximately fixes $\vec{\tau}_{\perp}$, while the inflow parameter $f_{\text{rad}}$ influences $f_{\text{r}}$.
Since both $f_{\text{r}}$ and $\vec{\tau}_{\perp}$ are fixed with the data, the parameters are $f_{\text{kep}}$ and $f_{\text{rad}}$ approximately linearly degenerate.
  \item Since G2 has a nearly radial orbit, inflow reduces the Mach number of the cloud relative to its background.
This significantly weakens our constraint on the strength of the magnetic field.
While \citetalias{McCourt2016} found a tight relationship between the field strength $\beta$ and by density $\alpha$, we find very low field strengths ($\beta \gg 1$) are allowed if the inflow velocity is well matched to G2's velocity ($f_{\text{rad}} \sim -1$).
This is exacerbated by the simplicity of our model, in which all velocities scale exactly $\propto \sqrt{G M_{\bullet}/r}$.
\end{enumerate}

\section{Kepler Elements}
\begin{figure*}
  \centering
  \includegraphics[width=\linewidth]{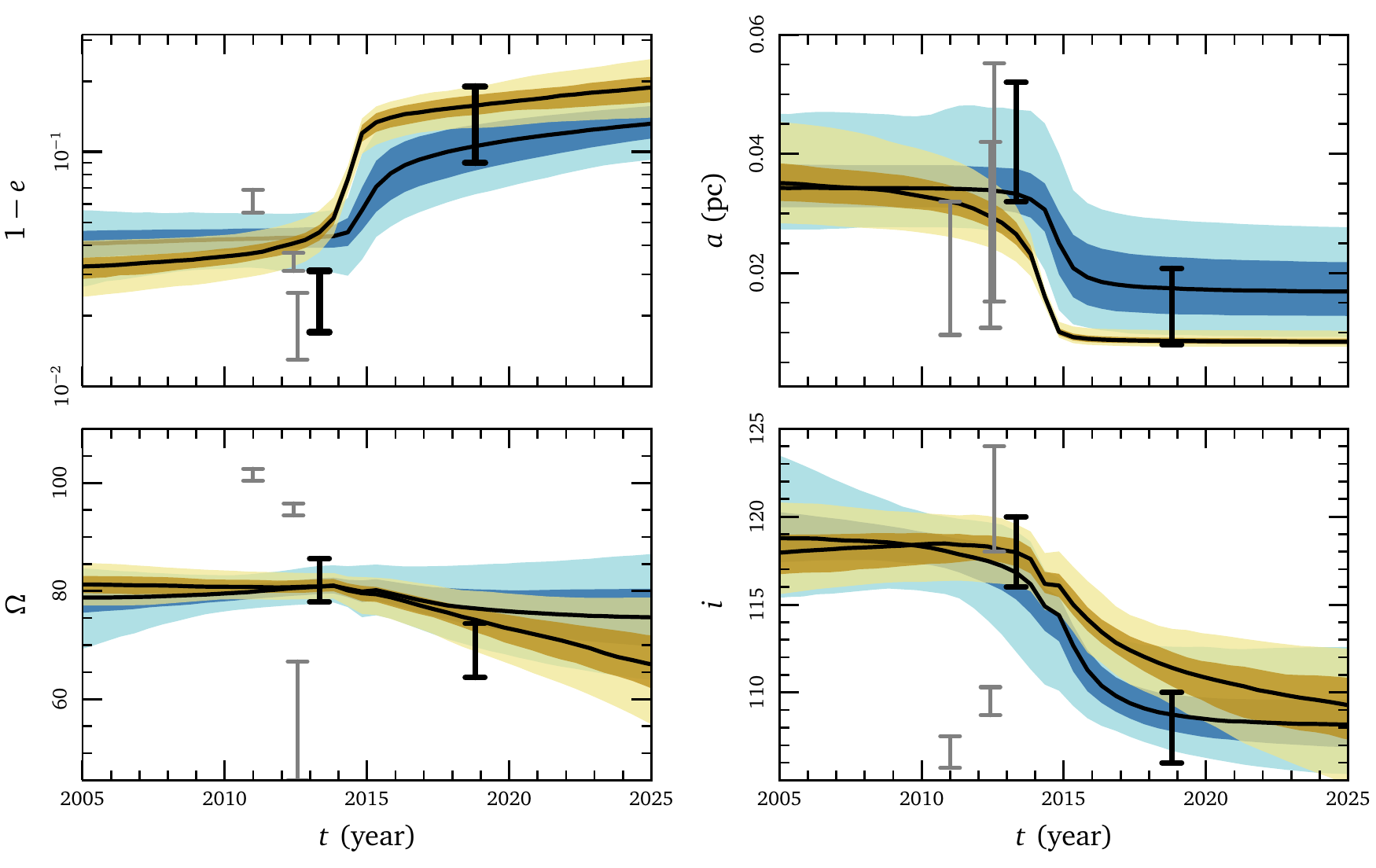}
  \caption{Time evolution of the Kepler elements in models with (blue) and without (yellow) inflow.  Shaded regions indicate 1- and 2-$\sigma$ quantiles.  Observational estimates for G1 and G2 are shown with error bars (G1 estimates are placed $12.8$ years in advance of G2's as we treat their orbits as a single trajectory).  In black we show those derived from the same Br-$\gamma$ dataset used in our fits \citep{Gillessen2013a,Gillessen2013b,Pfuhl2015}; in grey are those derived from $L^{\prime}$ data \citep{Phifer2013} data. We note that these data are slightly inconsistent. 
  The top panels show evolution in eccentricity $e$ and semi-major axis $a$. There is less energy evolution in models with inflow;  both $a$ and $e$ decrease less in these models than those without inflow. 
  The middle panels show the evolution in the orbital plane ($i$,$\Omega$).  Models with and without inflow evolve similarly as this is determined by the component of the force out of the orbital plane \fj.
  }\label{fig:kep-els}
\end{figure*}
In figure~\ref{fig:kep-els} we plot the time evolution of the Kepler elements.  Blue curves show the evolution in models with inflow; yellow curves show those without inflow from \citetalias{McCourt2016}.  
The top panels show eccentricity and semi-major axis of the best-fitting trajectories. 
Models with inflow result in less energy evolution before pericenter than in models without, because inflow reduces the relative velocity of the cloud with respect to the background. 
As the change in angular momentum is the same in both models, both semi-major axis and eccentricity evolve more slowly.  
The bottom panels show the change in orbital plane with time. Models with and without inflow show similar evolution. This is to be expected as changes in ($i,\Omega$) are driven by the component of the force out of the orbital plane \fj. Including a radial component \fr in our model does not directly affect this. In contrast, the evolution in the angle of pericenter $\omega$ is very different in models with and without inflow (see top panel of figure~\ref{fig:precession}). This is due to precession of the eccentricity vector. 

We also plot the derived Kepler elements for G1 and G2. 
In black we show those derived from the same Br-$\gamma$ dataset used in our fits \citep{Gillessen2013a,Gillessen2013b,Pfuhl2015}; in grey are those derived from $L^{\prime}$ data \citep{Phifer2013} data.  
We note that comparing Kepler elements is a very indirect comparison. If G1 and G2 are indeed gas clouds, their path is evolving on a timescale much less than an orbital period. Fitting Kepler elements to the data, while instructive, does not provide an accurate test for theoretical modeling. 

\end{document}